\documentclass[showpacs,preprintnumbers,amsmath,amssymb,twocolumn]{revtex4}
\usepackage{epsfig}
\usepackage{graphicx}
\usepackage{dcolumn}
\usepackage{bm}
\newcommand{\be}{\begin{equation}}
\newcommand{\ee}{\end{equation}}
\newcommand{\ba}{\begin{eqnarray}}
\newcommand{\ea}{\end{eqnarray}}

\begin{document}


\title{A new look at scalar mesons}
\author{L. Maiani}
\email{luciano.maiani@roma1.infn.it}
\affiliation{Universit\`{a} di Roma `La Sapienza' and I.N.F.N., Roma, Italy}
\author{F. Piccinini}
\email{fulvio.piccinini@pv.infn.it}
\affiliation{I.N.F.N. Sezione di Pavia and Dipartimento di Fisica Nucleare 
e Teorica, via A.~Bassi, 6, I-27100, Pavia, Italy}
\author{A.D. Polosa}
\email{antonio.polosa@cern.ch}
\affiliation{Centro Studi e Ricerche ``E. Fermi'', via Panisperna 89/A-00184
Roma, Italy}
\author{V. Riquer}
\email{veronica.riquer@cern.ch}
\affiliation{CERN Theory Department, CH-1211, Switzerland}


\begin{abstract}
Light scalar mesons are found to fit rather 
well a diquark-antidiquark description. 
The resulting nonet obeys mass formulae which respect,
to a good extent, the OZI 
rule. OZI allowed strong decays are 
reasonably reproduced by a single amplitude
describing the switch of a $q\bar q$ pair, which 
transforms the state into two colourless 
pseudoscalar mesons.
Predicted heavy states with one or more quarks 
replaced by charm or beauty are briefly described; 
they should  give rise to narrow states 
with exotic quantum numbers.\newline\newline
ROME1-1380/2004, FNT/T-2004/10, BA-TH/487/04, CERN-PH-TH/2004-124.

\end{abstract}

\maketitle

The well identified scalar mesons $a(980)$ ($I=1$, formerly called $\delta$) 
and $f(980)$ ($I=0$), have been frequently associated with 
$P$-wave, $q {\bar q}$ states~\cite{torn}. 
The main reason for this assignment is no doubt the fact that the other 
$P$-wave states, the axial and tensor nonets, are all well 
identified~\cite{pdg}. 
However, the $q\bar{q}$ assignment has never really worked in the 
scalar case. 
For one, $f$ is clearly associated to strange more than to up and down quarks, 
contrary to what the $I=0$ state degenerate to the $I=1$ one should do 
in a well-behaved $q\bar{q}$ nonet. Alternative identifications have 
been proposed in the past~\cite{close}, notably the $f$ as a bound 
$K\bar K$ molecule~\cite{isgur} or as a $(q)^2(\bar q)^2$ state~\cite{jaffe}.
Motivated by the recent discussion of exotic baryons as 
penta-quarks~\cite{jaffe-wilczek} and by 
the clear evidence by the KLOE Collaboration of a low mass~\cite{kloe} 
resonance, $\sigma(450)$, we examine in this paper the 
possibility that the lowest lying scalar mesons are $S-$wave bound states of 
a diquark-antidiquark pair. Following ref.~\cite{jaffe-wilczek} 
the diquark is taken to be in the fully antisymmetric combination of all 
quantum numbers, i.e. a colour anti-triplet, flavour anti-triplet, spin zero. 
The $(q)^2(\bar q)^2$  states make a flavour $SU(3)$ nonet. 
We propose to put the $\sigma$ in the remaining $I=S=0$ state, 
and to assign to the $S=\pm 1$ states the $\kappa(800)$,  a $K\pi$ 
resonance seen in several experiments, 
most recently in the $K \pi \pi$ spectrum from $D$ decays by the 
E791 Collaboration at FermiLab~\cite{e791}.
In addition to the quantum numbers, 
we consider the mass spectrum and the strong decays of the scalar mesons. 
A simple hypothesis on the way the $(q)^2(\bar q)^2$ states may transform 
into a pair of pseudo-scalar mesons is found to give a rather good, 
one parameter description of the decays allowed by 
the Okubo-Zweig-Iizuka et al. rule~\cite{ozi}. 
Addition of the remaining $SU(3)$ invariant couplings 
improves the description. 
In synthesis, we propose that scalar 
mesons below 1 GeV are diquark-antidiquark states. 
The $q\bar q$ $P-$wave scalar states, the partners of the tensor and 
axial nonet, will have 
to be found at higher masses. 
Some previous work in this direction can be found in the papers 
listed in~\cite{lista}.
We close the paper with a brief discussion 
of four-quark mesons with 
hidden and open charm or beauty, which should be characterized by narrow 
widths and spectacular decay modes.

{\bf {\sl Quantum numbers and mass formulae. }}
We denote by $[q_1q_2]$ the fully antisymmetric state of the two quarks 
$q_1$ and $q_2$. 
The composition of a few members of the nonet is as follows:    
\begin{eqnarray}
a^0(I=1,I_3=0) &=& \frac{1}{\sqrt{2}}\left( [s u] [\bar s \bar u] - 
[s d][\bar s \bar d] \right) \nonumber   \\
f_\circ(I=0) &=& \frac{1}{\sqrt{2}} \left( [s u] [\bar s \bar u] 
+ [s d] [\bar s \bar d] \right)  \nonumber \\
\sigma_\circ(I=0)&=& [u d] [\bar u \bar d] \nonumber  \\
\kappa &=& [u d] [\bar s \bar d], 
\end{eqnarray}
where:
\begin{eqnarray}
\vert f \rangle &=& \cos\phi \vert f_\circ \rangle +\sin\phi \vert \sigma_\circ \rangle \nonumber \\
\vert \sigma \rangle &=& -\sin\phi \vert f_\circ \rangle +\cos\phi \vert 
\sigma_\circ \rangle.
\label{eq:fsigmix}
\end{eqnarray}
The other members are easily reconstructed. For the neutrals, $I=0$, 
members we have introduced states with definite composition in strange quark 
pairs (exact OZI rule). 

Assuming octet symmetry breaking, masses depend on four parameters, 
according to the tensor expression (we use squared masses): 
\begin{equation}
M^2 = \frac{1}{2} \left\{ {\rm Tr}(S^2 m) 
+ \sqrt{3} c {\rm Tr}(S\lambda_8) {\rm Tr}S + \frac{3}{2} d 
[{\rm Tr}(S\lambda_0)]^2\right\}. 
\end{equation}
$S$ is the nonet 
scalar meson matrix, which we define according to: 
\begin{eqnarray}
S = 
\left( \begin{array}{ccc}
\frac{f_\circ + a^0}{\sqrt{2}} & a^+ & \kappa^+ \\
a^- & \frac{f_\circ - a^0}{\sqrt{2}} & \kappa^0 \\
\kappa^-  & \bar{\kappa^0}  &    \sigma_\circ
\end{array}\right), 
\end{eqnarray}
$m= {\rm diag}(\alpha,\alpha,\beta)$, with $\alpha$, $\beta$, $c$  and $d$ 
unknown coefficients, $\lambda_{0,8}$ are 
Gell-Mann's matrices and numerical coefficients have been introduced for convenience. 

In the limit $c=d=0$, the mass formula admits the 
states given above as mass eigenstates. In the more general case, for 
the $I\neq 0$ states we find (here and in the following, 
we indicate mass-squared with the particle's symbol): 
\begin{equation}
a=\alpha; \, \, \, \, \, \, \, \kappa = \frac{\alpha+\beta}{2}.
\end{equation}
The mass-squared matrix of $I=0$ states is: 
\begin{eqnarray}
\mu^2 = 
\left( \begin{array}{cc}
\alpha + 2(c+d) & \frac{1}{\sqrt{2}}(-c + 2d) \\
\frac{1}{\sqrt{2}}(-c+2d) & \beta - 2c + d \\
\end{array}\right).
\end{eqnarray}
We can eliminate $c$ and $d$ in favor of physical masses and $f-\sigma$ 
mixing~(\ref{eq:fsigmix}). 
We remain with one overall relation which fixes the 
$f-\sigma$ mixing angle as 
function of the masses. 
Taking, for simplicity, $f(980)$ 
degenerate with $a(980)$ we find: 
\begin{equation}
\cos2\phi + 2 \sqrt{2} \sin2\phi = 1+4\frac{a +\sigma - 2\kappa}{a-\sigma}.
\end{equation}
For the masses we take the values reported in Table~I. 
The $\sigma$ mass is known with large errors. From the above 
equation we find:
\begin{eqnarray}
\begin{array}{cc}
\tan2\phi=-0.07, & \; \; \; \sigma = (570\; {\rm MeV})^2 \\
\tan2\phi=-0.19, & \; \; \; \sigma = (470\; {\rm MeV})^2 \\
\tan2\phi=-0.31, & \; \; \; \sigma = (370\; {\rm MeV})^2. \\
\end{array}
\end{eqnarray}
Since we are holding $f$ degenerate with $a$, the $\sigma$ 
mass is pushed down as mixing becomes more negative. 
The $\sigma$ mass-squared gets to zero for $\tan2\phi=-0.48$, which gives the 
lowest bound to the mixing angle.
Mixing is small because the OZI rule is respected in the
physical mass spectrum. The spectrum is inverted with respect to 
$q\bar q$ nonets: the isolated $I=0$ state is the lightest one and strange 
particles come next. This is a most evident indication in favour of the 
four-quark nature of the scalar nonet.	
\begin{table}
\begin{center}
\begin{tabular}{|l|l|l|l|l|}
\hline
Meson & Mass (MeV) & Source \\  \hline
$\sigma$      & $478 \pm 24 \pm 17$          & \cite{kloe} \\
\hline
$\kappa$      & $797 \pm 19 \pm 43$          & E791~\cite{e791} \\
\hline
$f$           & $980 \pm 10$               & PDG~\cite{pdg}\\
\hline
$a$           & $984.7 \pm 1.2$            & PDG~\cite{pdg}\\
\hline
\end{tabular}
\end{center}
\caption{ Experimental values for the scalar meson masses.}
\end{table}
\newline
{\bf {\sl Strong decays.}} 
Diquarks being colour antitriplets, they cannot be separated 
by their antiparticles. As soon as the distance between the members of the 
pair gets large enough, a $q-\bar q$ pair is created out of the vacuum and the 
pair dissociates into a baryon-antibaryon. This process cannot take place 
spontaneously, however, the $S-$wave scalar mesons are quite below threshold 
for the baryon-antibaryon decay.  An alternative mechanism is that a 
quark-antiquark pair is switched between the members of the pair, to form a 
pair of colourless $q-\bar q$ states, which can indefinitely separate 
from each other, see Fig. 1. The lightest decay channel is a pair of 
pseudoscalar mesons. In the exact $SU(3)$ limit there is only one amplitude, 
$A$, to describe this process. The amplitude for the switch is not expected to 
have any particular suppression for the $S-$wave scalars, since 
there is no barrier for the diquark and the antidiquark to overlap. This is 
at variance with the case of the two diquarks in the exotic baryons, which 
avoid getting close to each other due to Pauli blocking~\cite{jaffe-wilczek}. 
\begin{figure}[ht]
\begin{center}
\epsfig{
height=3truecm, width=7.5truecm,
        figure=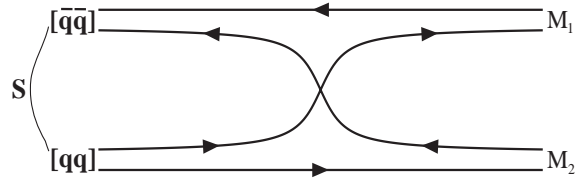}
\caption{\label{fig.1} \footnotesize 
The decay of a scalar meson $S$ made up of 
a diquark-antidiquark pair
in two mesons $M_1 M_2$ made up of standard $(q \bar{q})$ pairs.
}
\end{center}
\end{figure}
Our picture has some connection with baryonium states~\cite{rossi-veneziano}) and with the 
$K\bar K$ molecule picture~\cite{close}. 
In the latter case, however, the analogy is only superficial. The meson states we 
are considering correspond to quite 
different configurations than a $K-\bar K$ molecule. Indeed, they are completely 
orthogonal to them. The amplitude $A$ describes the tunneling from the bound 
diquark pair configuration to the meson-meson pair, made by the unbound, 
final state particles. With the aid of Fig.~1, the amplitudes for different 
decays are easily computed. For instance, we have: 
\begin{eqnarray}
[s u]_{\bar{3}_c}[\bar s \bar d ]_{3_c} &\to& (s \bar d)_{1_c} ({\bar s} u)_{1_c} - 
(s \bar s)_{1_c} ({\bar d} u)_{1_c} \nonumber\\
&=& {\bar K^0} K^+ - \pi^+ \eta_s.
\end{eqnarray}
For convenience we introduce the combinations: 
\begin{equation}
\eta_q = \frac{({\bar u} \gamma_5 u + {\bar d} \gamma_5 d )}{\sqrt{2}}; \; \; \; \eta_s = 
{\bar s}\gamma_5 s, 
\end{equation}
which can be expressed in terms of the physical $\eta$ and $\eta'$ fields and of 
the pseudoscalar meson mixing angle 
(we use mass-squared formulae and correspondingly $\sin\phi_{PS}=0.19$).
After that, we find: 
\begin{eqnarray}
& &{\rm Ampl.}(a^+ \to {\bar K^0} K^+) = A; \nonumber \\
& &{\rm Ampl.}(a^+ \to \pi^+ \eta) = A
\left( -\sqrt{\frac{2}{3}} \cos \phi_{PS} 
+ \sqrt{\frac{1}{3}} \sin \phi_{PS}\right) \nonumber \\
& &\simeq -0.69 A.
\end{eqnarray}

For the relevant decays, we give the result in the form of 
an effective Lagrangian and in terms of the unmixed fields 
$f_\circ$ and $\sigma_\circ$. 
In this form, the amplitude $A$ has the dimension of a mass. 
\begin{eqnarray}
{\cal L} =  & &A \left[  f_\circ \left( -\frac{{\bar K} K}{\sqrt{2}} 
+ \eta_q \eta_s \right) 
- \sigma_\circ \left( \frac{{\bf \pi} \cdot {\bf \pi}}{2} 
+ \frac{\eta_q^2}{2} \right) \right. \nonumber \\
& & \left. + a^0 \left( \frac{{\bar K} \tau_3 K}{\sqrt{2}} - \pi^0 \eta_s\right)  
+ \left( \frac{\bar{K^+} \pi^0}{\sqrt{2}} + {\bar K^0}\pi^- \right)\kappa^+ \right. \nonumber \\
& & \left. + \ldots \right]. 
\end{eqnarray}
Decay rates are expressed as:  
\begin{equation}
\Gamma(S\to i) = \frac{A^2}{8 \pi} \frac{p}{M_s^2} x_{s\to i}, 
\label{eq:gammsi}
\end{equation}
where $p$ is the decay momentum, $M$ the mass of the 
scalar meson and $x_{s \to i}$ a factor which includes numerical coefficients in 
the individual amplitudes and isospin multiplicities. 
Without attempting a systematic fit, we take for $A$ the value: 
$A= 2.6$~GeV 
and give in Table~II the 
corresponding calculated rates, compared to the available experimental 
information.  For simplicity, statistical and systematic errors in the 
experimental values have been combined in quadrature. Some comments are 
in order. 

1.~We have taken from ref.~\cite{barberis} the total 
width
$\Gamma_{tot} (a_0)= 72\pm16$~MeV
and the $K\bar K$ branching ratio
${\cal B}(a_0\to K\bar K)= 0.17 \pm 0.03$ 
thus obtaining:
\begin{eqnarray}
&&\Gamma(a_0\to \eta \pi)= 60 \pm 13 \;\; {\rm MeV}, \\
&&\Gamma(a_0\to K \bar K) = 12 \pm 3 \; \; {\rm MeV}.
\end{eqnarray}

2.~We compute the decay momentum with the central values of the parent mass, 
with the exception of the decay $a\to K \bar K$, which is below threshold at 
the central mass value. In this case we have averaged the decay momentum 
over a Breit-Wigner, using the  $\Gamma_{tot} (a)$ given above, and 
find:
$\langle p(a\to K \bar K )\rangle \approx 84$~{\rm MeV}, 
\begin{table}
\begin{center}
\begin{tabular}{|c|c|c|c|c|}
\hline
  &  \multicolumn{2}{c|}{$\pi \pi$} &  \multicolumn{2}{c|}{$K \bar K$}   \\  
\hline 
 $\sigma$ & 345~MeV & $324 \pm 50$~MeV & \multicolumn{2}{c|}{-} \\
\hline
$f$ & $g_\pi < 0.02$ & $g_\pi=0.19 \pm 0.05 $ & $g_K=0.28$ & $g_K = 0.40 \pm 0.6$  \\
\hline
& \multicolumn{2}{c|}{$\eta \pi$} & \multicolumn{2}{c|}{} \\
\hline
$a$ & 43~MeV & $60 \pm 13$~MeV & 23~MeV & $12 \pm 3$~MeV \\
\hline
 & \multicolumn{2}{c|}{$K \pi$} & \multicolumn{2}{c|}{} \\
\hline
$\kappa$ & 138~MeV & $410 \pm 100$~MeV & \multicolumn{2}{c|}{-} \\
\hline
\end{tabular}
\end{center}
\caption{
Fit with a single 
parameter $A=2.6$~GeV. For $g_\pi$ we have reported the upper 
limit to the decay rate obtained from the $f-\sigma$ 
mixing considered previously, 
see text.}
\end{table}
which gives the value of the partial width reported in Table~II.

3.~In the case of $f \to K \bar K$  or $\pi \pi$, the authors of 
ref.~\cite{barberis}  define: 
\begin{equation}
\Gamma(S\to i) = g_i p(M)
\end{equation}
and fit the data to a Breit-Wigner formula with mass-dependent 
width, thus giving directly the values of $g_i$ that we report 
in the Table~II. 

It is interesting to see if the agreement can be improved by introducing 
other $SU(3)$ allowed couplings. In the exact $SU(3)$ limit there are four 
couplings, but one refers to a pure singlet-to-singlets amplitude, which 
is not relevant to the above decays. Restricting to the other three 
couplings, we write the effective Lagrangian according to: 
\begin{eqnarray}
{\cal L} &=&  (S_i^j) \epsilon_{jlm} \epsilon^{ikn} [a M_k^l M_n^m + b \delta_k^l (M^2)_n^m+ \nonumber \\
&+& c \delta_k^l (M)_n^m {\rm Tr}M], 
\end{eqnarray}
$M$ represents the nonet pseudoscalar matrix, analogous to $S$, and we have made explicit 
the four quark nature of the scalar nonet. 
The first coupling corresponds to the switch amplitude, Fig.~1. The other two couplings correspond to 
amplitudes where one pair annihilates into a flavour singlet (gluons) that 
transforms into a $q \bar q$ flavour singlet pair, violating the OZI rule. 
For $a=A$, $b=c=0$ we reproduce the previous results. We obtain the effective 
Lagrangian:  
\begin{eqnarray}
{\cal L} &=& f_\circ \left[ b \sqrt{2} \frac{\pi \cdot \pi}{2} - (a-3b)\frac{{\bar K} K}{\sqrt{2}}+\ldots \right] 
\nonumber \\ 
&+& \sigma_\circ \left[ -(a-2b) \frac{\pi \cdot \pi}{2} + b {\bar K} K + \ldots \right] \nonumber \\
&+& a^0 \left[ (a-b) \frac{{\bar K}\tau_3 K}{\sqrt{2}} - (a-c) \eta_s \pi^0 \right.\nonumber \\
&-& \left. \sqrt{2} (b-c) \eta_q \pi^0 + \ldots \right] \nonumber \\
&+& (a-b) \left( \frac{\bar{K^+}\pi^0}{\sqrt{2}} + {\bar K^0}\pi^-\right) \kappa^+ + \ldots 
\end{eqnarray}
The amplitude for $a\to \eta \pi$ receives a new contribution from $c$ and 
is now independent from the others. The three OZI allowed amplitudes 
$\sigma \to \pi \pi$, $a_0/f_0 \to K \bar K$ are now predicted to be linearly spaced with $b$. 
From the experimental values in Table~II we find:
\begin{eqnarray}
& &\vert a - 2b \vert = 2.6\;\; {\rm GeV} \; (=A)\; ({\rm from}\; \; \sigma \to \pi \pi), \nonumber \\
& &\vert a-3b \vert = 3.1\;\; {\rm GeV} \; ({\rm from}\; \; f_0 \to  K \bar K), \nonumber \\
& &\vert a-b \vert = 1.8\;\; {\rm GeV} \; ({\rm from}\; \;  a_0 \to K \bar K), 
\end{eqnarray} 
which are indeed equally spaced with: 
$b= - 0.7$~GeV. 
We find further: 
$\Gamma(\kappa)= 66\;\;{\rm MeV}; \; \; \; g_\pi = 0.06.$
The $c$ (annihilation) coupling should be small: with $c$ exactly zero 
we get $\Gamma(a\to \eta\pi)=30$~MeV; cfr. Table~II.
A better agreement is found with data for the OZI allowed channels, 
except for the $\kappa$ width, which is too low (but also known with 
large uncertainties). The large value of $A$ seems indicative of a short 
distance effect, making perhaps more justifiable the use 
of flavour $SU(3)$ symmetry. These results reinforce considerably the case of the 
scalar mesons as $(q)^2(\bar q)^2$ states. 
A notable exception is the OZI forbidden decay $f\to \pi \pi$, which turns 
out to be too small, even allowing for the full $SU(3)$ couplings. 
It is quite possible that this 
decay proceeds via a different mechanism. One possibility we would like 
to suggest is: 
\begin{equation}
f \to K \bar K \to \pi \pi, \nonumber
\end{equation}
with the first step via off-shell $K\bar K$ states and the 
second by a strong, OZI allowed, process. A calculation of this effect, 
with the second step mediated by $K^*$ and $\kappa$ exchange is under 
consideration. It is a calculation that closely resembles those 
performed in the $K \bar K $ molecule picture.

{\bf {\sl Open and hidden charm scalar mesons.}} A firm prediction of the 
present scheme is the existence of analogous states where one or more 
quarks are replaced by charm or beauty. We consider the case of charm, 
extension to beauty is obvious. Open charm scalar mesons of the form
$S=[cq][\bar{q}\bar{q}]$, 
fall into characteristic ${\bf 6}\oplus \bar{{\bf 3}}$ multiplets of
$SU(3)_f$. The $\bar{\bf 3}$ has the same conserved quantum numbers
of $c\bar q$ states, but the ${\bf 6}$ contains exotic states which
should be very conspicuous. 

Open charm states are classified as follows.
\\$S=1$:
\begin{eqnarray}
a_{c \bar{s}}(I=1), f_{c\bar{s}}(I=0)=[cq][\bar q \bar s]. \nonumber
\end{eqnarray}
They form a degenerate triplet-singlet similar to the $a/f$ 
complex, but with charges $0,+1,+2$. OZI allowed decays are:
\begin{eqnarray}
&&a_{c \bar s}\to D_s \pi, DK \; \; \;  ({\rm E_{thr}} = 2103.6,\; 2367\; {\rm MeV}), \nonumber \\
&&f_{c \bar s}\to D_s\eta, DK \; \; \;  ({\rm E_{thr}} = 2515.9,\; 2367\; {\rm MeV}). \nonumber
\end{eqnarray}
$S=0$:
\begin{eqnarray}
&&\delta_{c}(I=1/2)=[cs][\bar q \bar s],   \nonumber \\
&&S_c(I=1/2)=[cq][\bar u \bar d]. \nonumber
\end{eqnarray}
The two isodoublets are superpositions of ${\bf 6}$ and 
$\bar{\bf 3}$ components with decays:
\begin{eqnarray}
&&\delta_c\to D\eta, D_s \bar K \; \; \;  ({\rm E_{thr}} = 2416.6,\; 2466.3\; {\rm MeV}), \nonumber \\
&&S_c\to D\pi \; \; \;  ({\rm E_{thr}} = 2004.3\; {\rm MeV}). \nonumber
\end{eqnarray}
$S=-1$:
\begin{eqnarray}
&&\omega_{c}(I=0)=[cs][\bar u \bar d],  \nonumber\\
&&\omega_c \to D\bar K \; \; \;  ({\rm E_{thr}} = 2367\; {\rm MeV}). \nonumber
\end{eqnarray}
Hidden charm states of the form: $[cq][\bar c \bar q]$ fall into
${\bf 8}\oplus {\bf 1}$ multiplets of $SU(3)$ again producing very 
exotic states. We simply mention the degenerate 
isotriplet-isosinglet complex: $a_{c\bar c}(I=1)$, $f_{c\bar c}(I=0)$, 
equal to the $a/f$ complex with $s$ replaced by $c$, with 
characteristic decays:
\begin{eqnarray}
& & a_{c \bar c} \to \eta_c \pi, \; \; D \bar D  \; \; \; ({\rm E_{thr}} 
= 3114.7,\; 3738.6\; {\rm MeV}), \nonumber  \\
& & f_{c \bar c} \to \eta_c \eta, \; \; D \bar D \; \; \; ({\rm E_{thr}} 
= 3527,\; 3738.6\; {\rm MeV}). \nonumber  
\end{eqnarray}
We expect quark pair annihilation to be suppressed by asymptotic freedom. 
Thus the decay rates into exclusive channels should be well described by the 
simple switch amplitude (Fig. 1).
By scaling from Eq.~(\ref{eq:gammsi}) one finds widths~$\approx O(10)$~MeV.

Narrow states decaying into open or hidden charm states and a pseudoscalar 
meson are being discovered at PEPII and BELLE and in fixed target experiments 
(FNAL). Analysis of these states in term of four quark states has been done 
in some cases~\cite{terasaki}. Further experimental search for exotic states 
is crucial.

{\bf {\sl Acknowledgments}}. 
We thank P. Franzini and G. Isidori for useful information on KLOE results. 
FP whishes to thank partial support from CERN where
part of this work has been made. ADP thanks the Physics Department of the
University of Bari for its kind hospitality.


\begin{thebibliography}{99}
\bibitem{torn} N.A.~T\"{o}rnqvist, Z. Phys. {\bf C68} (1995) 647.
\bibitem{pdg} S.~Eidelman et al., Phys. Lett. {\bf B592} (2004) 1.
\bibitem{close} F.E.~Close and N.A.~Tornqvist, J.~Phys.~{\bf G28} (2002) R249 
and references therein.
\bibitem{isgur} J.~Weinstein and N.~Isgur, Phys. Rev. Lett. {\bf 48}(1982) 659.
\bibitem{jaffe} R.L.~Jaffe, Phys. Rev. {\bf D15} (1977) 281; 
R.L.~Jaffe and F.E.~Low, Phys. Rev. {\bf D19} (1979) 2105; 
M.~Alford and R.L.~Jaffe, Nucl. Phys. {\bf B578} (2000) 367.
\bibitem{jaffe-wilczek}
S.~Nussinov, hep-ph/0307357; M.~Karliner and H.J.~Lipkin, hep-ph/0307243. 
R.L.~Jaffe and F.~Wilczek, Phys. Rev. Lett. {\bf 91} (2003) 232003.
\bibitem{kloe} KLOE Collaboration (A. Aloisio {\it et al.}), Phys. Lett. 
 {\bf B537} (2002) 21; E.M.~Aitala {\it et al.}, Phys.~Rev.~Lett. 
{\bf 86} (2001) 770.
\bibitem{e791} E.M.~Aitala {\it et al.}, Phys. Rev. Lett. {\bf 89} (2002) 
121801.
\bibitem{ozi} G.~Zweig, CERN report S419/TH412 (1964), unpublished;
S.~Okubo, Phys. Lett, {\bf 5}(1963) 165;
J.~Iizuka, K.~Okada and O. Shito, Prog. Theor. Phys. {\bf 35} (1966) 1061.
\bibitem{lista} H.~Lipkin, Phys. Lett. {\bf B70}, 113 (1977); D.~Black, 
A.H.~Fariborz, F.~Sannino and J. Schechter, Phys. Rev. {\bf D59}, 
074026 (1999) where, unlike this paper, derivative couplings were used;
J.R. Pelaez, Phys. Rev. Lett. {\bf 92}, 102001 (2004); M.~Alford 
and R.L.~Jaffe, Nucl.Phys. {\bf B578}, 367 (2000) for lattice studies; 
A nice review with a complete list of reference 
can be found in R.L.~Jaffe, {\it Exotica}, hep-ph/0409065. 
\bibitem{rossi-veneziano} G.C.~Rossi and G.~Veneziano,
Nucl. Phys. {\bf B123} (1977) 507; Phys. Lett. {\bf B70}, 255 (1977);
for a recent update see hep-th/0404262.
\bibitem{barberis} D.~Barberis et al., Phys. Lett. {\bf B453} (1999) 325.
\bibitem{terasaki} K.~Terasaki, 
{\sl ``Charmed Scalar Mesons and Related''}, hep-ph/0405146
\end{thebibliography}
\end{document}